\documentclass[aps, reprint,10.0pt,twocolumn,superscriptaddress]{revtex4-2}
\usepackage{amsmath}
\usepackage{latexsym}
\usepackage{amssymb}
\usepackage{graphics,epstopdf}
\usepackage{hyperref}
\usepackage{soul}
\usepackage{braket}
\hypersetup{
    colorlinks = true,
    linkcolor =blue,
	citecolor=blue, 
	urlcolor=blue 
}
\usepackage{cancel}

\newcommand{\ed}{\end{document}}
\newcommand{\beq}{\begin{equation}}
\newcommand{\eeq}{\end{equation}}

\usepackage{mathtools}
\usepackage{natbib}

\newtheorem{lemma}{Lemma}
\newtheorem{theorem}[lemma]{Theorem}
\newtheorem{definition}[lemma]{Definition}

\begin{document}
\title{Effect of order of transfer matrix exceptional points on transport at band edges}

\author{Madhumita Saha}
\email{madhumita.saha@icts.res.in} 
\affiliation{International Centre for Theoretical Sciences, Tata Institute of Fundamental Research,
Bangalore 560089, India}

\author{Bijay Kumar Agarwalla}
\email{bijay@iiserpune.ac.in}
\affiliation{Department of Physics, Indian Institute of Science Education and Research Pune, Dr. Homi Bhabha Road, Ward No. 8, NCL Colony, Pashan, Pune, Maharashtra 411008, India}

\author{Manas Kulkarni}
\email{manas.kulkarni@icts.res.in} 
\affiliation{International Centre for Theoretical Sciences, Tata Institute of Fundamental Research,
Bangalore 560089, India}

\author{Archak Purkayastha}
\email{archak.p@phy.iith.ac.in}
\affiliation{Department of Physics, Indian Institute of Technology, Hyderabad 502284, India}

\date{\today} 
\begin{abstract}
{Recently, it has been shown that, in one dimensional fermionic systems, close to band edges, the zero temperature conductance  scales as $1/N^2$, where $N$ is the system length. This universal subdiffusive scaling of conductance at band edges has been tied to an exceptional point (EP) of the transfer matrix of the system that occur at every band edge. Further, in presence of bulk dephasing probes, this EP has been shown to lead to a counterintuitive superballistic scaling of conductance, where the conductance increases with $N$ over a finite but large regime of system lengths. In this work, we explore how these behaviors are affected by the order of the transfer matrix EP at the band edge. We consider a one-dimensional fermionic lattice chain with a finite range of hopping. Depending on the range of hopping and the hopping parameters, this system can feature band edges which correspond to arbitrarily higher order EPs of the associated transfer matrix. Using this system we establish in generality that, in absence of bulk dephasing, surprisingly, the universal $1/N^2$ scaling of conductance is completely unaffected by the order of the EP. This is despite the fact that existence of transfer matrix EP is crucial for such behavior. In presence of bulk dephasing, however, the phase coherence length, the extent of the superballisitic scaling regime and the exponent of superballistic scaling, all encode the order of the transfer matrix EP.}
\end{abstract}

\maketitle


\section{Introduction}
Non-Hermitian physics, which deals with how properties of non-Hermitian matrices occur in the description of various physical systems, has evolved into an extremely active research area \cite{non-hermitian-review}. A crucial distinction of non-Hermitian matrices from Hermitian counterparts is that they may not be diagonalizable. Parameter points where the non-Hermitian matrices are non-diagonalizable are called the exceptional points (EPs). An EP occurs when several eigenvalues and eigenvectors of the matrix coalesce. The number of such coalesced eigenvalues or eigenvectors gives the order of the EP. Most initial studies focused on EPs of second order, while more recently, higher order EPs have been of significant interest \cite{higher_order_exceptional,open_exceptional,toplology_non_hermitian,EPD_photonic,wave_guideEPD,symmetry_higher_order_EP,higher_order_review}. It has been established that EPs are of relevance to various device applications such as sensing \cite{sensing_EPD,sensing1,sensing_review,sensing_review1}, lasing \cite{band_edge_EP_transfer_matrix}, control of  spontaneous emission \cite{spontaneous_emission}, and perfect absorption \cite{perfect_absorber}. An important aspect of EPs which governs many of these applications is that in the vicinity of the EPs, the system is sensitive towards perturbation, with higher-order EPs being more sensitive \cite{}.

Most studies consider EPs of an effective non-Hermitian Hamiltonian, i.e, the matrix that directly governs dynamics of the system \cite{non-hermitian-review}. Thus, the dynamics of the system is non-unitary. In this connection, there has been significant recent interest in understanding how non-Hermitian Hamiltonians and their EPs affect the transport properties of a system \cite{non-hermitian-transport1,non-hermitian-transport2,non-hermitian-transport3,chiral_EP}. In contrast, even in systems governed by Hermitian Hamiltonians (unitary dynamics), non-Hermitian matrices can occur and their EPs can affect the physics non-trivially \cite{Wang_2019, Lee_2024, Sergeev_2023, universal_subdiffusive, superballistic1, finite-range-subdiffusive, hu_2024}. For example, the transfer matrix, which occurs in description of lattice systems, is inherently non-Hermitian. It does not directly govern the dynamics of the system, but rather governs the energy dispersion and transport properties \cite{Molinari_1997, Molinari_1998, transfer_matrix1, transfer_matrix2, universal_subdiffusive,finite-range-subdiffusive,superballistic1}. It has been recently shown that in ordered nearest neighbour one dimensional systems, band edges correspond to second order EPs of the transfer matrix \cite{universal_subdiffusive}. As a consequence, when chemical potential is at any band edge, and the system is connected to two baths at two ends, conductance $\mathcal{G}(\mu)$ shows a universal scaling, $\mathcal{G}(\mu)\sim N^{-2}$, where $N$ is system length \cite{universal_subdiffusive}. This occurs when transport is coherent. In presence of small incoherent effects (dephasing) due to a surrounding environment in the bulk, there occurs a counterintuitive superballistic scaling of conductance, where conductance increases with $N$ as a power-law  over a finite but large regime of system lengths \cite{superballistic1}. These anomalous transport behaviors are of particular interest since they occur in absence of any disorder, and have no analog in isolated systems, i.e, in absence of the baths. The subdiffusive scaling of conductance at band edge have also been  explored in systems with next-nearest neighbour hopping \cite{finite-range-subdiffusive} where it is also shown to stem from transfer matrix EP, as well as in two-dimensional systems \cite{junaid_subdiffusive} and power-law decaying hopping systems {\cite{subdiffusive_long_range}, where the connection to any transfer matrix EP is not obvious.

In this work, we investigate the effect of the order of the transfer matrix EP on the above-mentioned counterintuitive transport behavior at the band-edges. To do so, we consider coherent and incoherent transport through a fermionic chain with finite-range hopping. The range of hopping is $n$, which is arbitrary, but does not scale with the system length $N$. The associated transfer matrix of the system is of size $2n \times 2n$, and can feature band edges corresponding to transfer matrix EPs of order up to $2n$. In fact, there is a systematic way to generate such higher order EPs of the transfer matrix in such cases \cite{isolated_draft}.  Using this system we establish analytically and demonstrate numerically that the subdiffusive scaling of conductance in the coherent case is completely unaffected by the order of the transfer matrix EP. That is, $\mathcal{G}(\mu)\sim N^{-2}$ when $\mu$ equals a band edge, irrespective of the order of the transfer matrix EP. This is surprising because it is also clear that existence of transfer matrix EP is crucial to the subdiffusive scaling. On the contrary, the superballistic scaling in presence of bulk dephasing modelled via B{\"u}ttiker voltage probes is strongly affected by the order of the transfer matrix EPs. In particular, the exponent of superballistic scaling encodes the order of the transfer matrix EP. The behavior of phase coherence length with the strength of dephasing, which, in turn, governs the extent of the superballistic scaling regime, also depends on the order of the tranfer matrix EP.

The paper is organized as follows: In Sec.~\eqref{isolated_transfer}, we introduce the isolated lattice setup and discuss some relevant properties of the finite-range hopping model. In Sec.~\eqref{open}, we discuss the open version of this setup by introducing the boundary reservoirs, one at each end. In Sec.~\eqref{probe_absence} we first discuss the results in the absence of BVPs and investigate the two-terminal conductance scaling with system size in the presence of higher order EPs. In Sec.~\eqref{with_probes}, we discuss the conductance scaling in the presence of BVPs and report the sensitivity of higher order EPs in conductance scaling. 
In Sec.~\eqref{sec:summary}, we summarize our results along with an outlook.  Certain details are delegated to the appendices.

\begin{figure}
\includegraphics[width=\columnwidth]{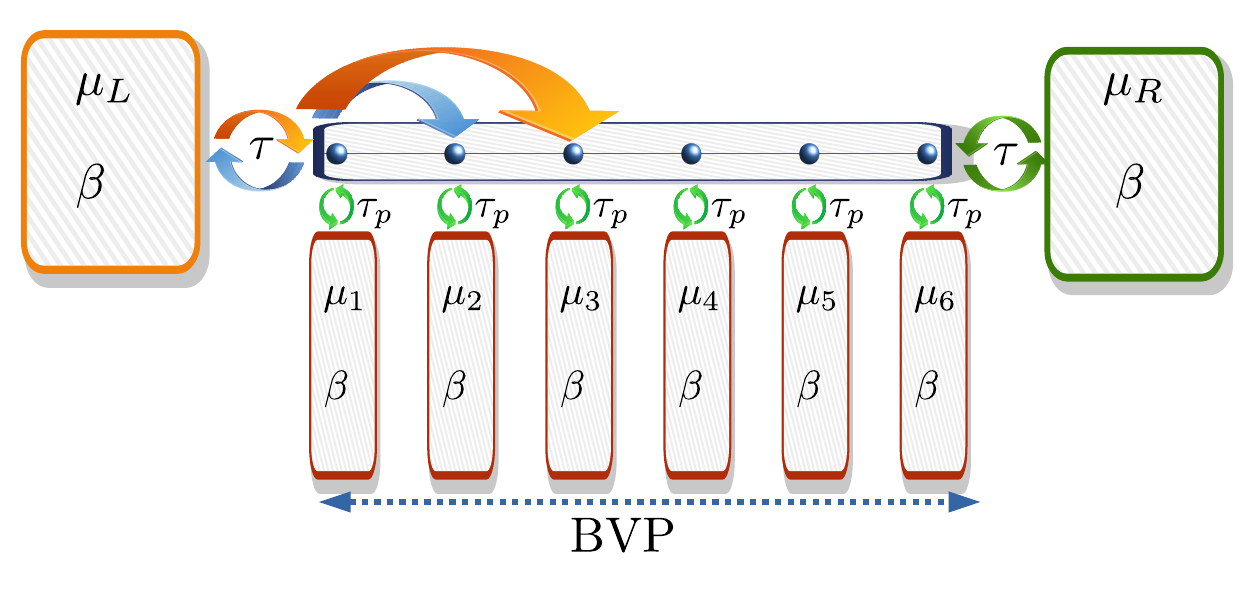} 
\caption{Schematic of the model we consider. Our setup represents a tight-binding lattice chain that is connected to left bath (L), right bath (R), and BVPs, all of which are kept at zero temperature. The coupling strength between the system and the left/right bath is given by $\tau$, while between the system and the B{\"u}ttiker probe is denoted by $\tau_P$. The chemical potential of the probes $\mu_n, n=1,2, \cdots N$ are determined by demanding the zero particle current between $n$-th site and $n$-th probe. In this work, we consider two different scenarios for such open quantum setup: (i) in absence of probes ($\tau \neq 0, \tau_P=0$) in Sec.~\eqref{probe_absence}, and (ii) in presence of probes ($\tau \neq 0, \tau_P \neq 0$) in Sec.~\eqref{with_probes}.}
\label{fig:schematic} 
\end{figure}

\section{Isolated lattice setup and relevant properties}
\label{isolated_transfer}
In this section,  we first discuss the isolated system of a fermionic one-dimensional lattice with finite-range hopping $n$, provide the transfer matrix of the lattice and then discuss the EPs of the transfer matrix. 
\subsection{Hamiltonian and the transfer matrix}
The Hamiltonian of the lattice chain can be written as
\begin{align}
\hat{H}_C=-\sum_{m=1}^n \sum_{i=-N/2+1}^{N/2} t_m \hat{c}^{\dagger}_i \hat{c}_{i+m} + h.c.,
\label{lattice-hamiltonian}
\end{align}
where, $\hat{c}^{\dagger}_i (\hat{c}_i)$ is the fermionic creation (annihilation) operator, and $h.c.$ stands for Hermitian conjugate. Here, $t_m$ is the hopping strength when the particle hops from $i-$th site to $(i+m)-$th site, $N$ is the system size which we considered to be even. The subscript $c$ in $\hat{H}_C$ refers to the lattice chain. The Hamiltonian can be cast in the form

\begin{align}
\label{single_particle_def}
\hat{H}_C=\sum_{\ell,m=-N/2+1}^{N/2} \bra{\ell}\mathbf{H}\ket{m} \hat{c}_{\ell}^\dagger \hat{c}_m,
\end{align}
where $\bra{\ell}\mathbf{H}\ket{m}$ are components of a $N \times N$ symmetric matrix such that only the components allowed by Eq.~\eqref{lattice-hamiltonian} are non-zero. 
The eigenvalue equation for the single particle matrix $\mathbf{H}$ is given by 
\begin{align}
 \mathbf{H}\ket{\psi}=\omega\ket{\psi}, ~~\ket{\psi}=\sum_{\ell=-N/2+1}^{N/2}  \psi_\ell \ket{\ell},  
 \label{Hpsi}
\end{align}
where $\omega$ is the single particle eigenvalue and $\ket{\psi}$ is a vector whose $\ell$th element, $\psi_\ell$, gives the component of the corresponding single-particle eigenvector at site $\ell$. Following Eqs.~\eqref{lattice-hamiltonian} and \eqref{Hpsi}, for a given $\omega$, components of the single particle eigenvector $\ket{\psi_{\ell}}$ can be obtained using the recursion relation
\begin{align}
\label{recursion}
\ket{\psi_{\ell}^n} = \mathbf{T}(\omega) \ket{\psi_{\ell-1}^n},
\end{align}
where $\ket{\psi_{\ell}^n}$ is a $2n \times 1$ column vector given as,
\begin{align}
\label{psi_l_n_def}
\ket{\psi_{\ell}^n} =\begin{pmatrix} \psi_{\ell+n} \\ \psi_{\ell+n-1}\\ \vdots \\ \psi_{\ell-n+3}\\ \psi_{\ell-n+2}\\ \psi_{\ell-n+1}   \end{pmatrix}
\end{align}
and the transfer matrix $\mathbf{T}(\omega)$ \cite{matrix_inversion,finite-range-subdiffusive} is given as,
\begin{align}
\label{transfermat1}
&\mathbf{T}(\omega)=\nonumber \\
&\begin{pmatrix} -\frac{t_{n-1}}{t_n} & -\frac{t_{n-2}}{t_n}& \ldots & -\frac{\omega}{t_n} & \ldots & -\frac{t_{n-2}}{t_n} & -\frac{t_{n-1}}{t_n} & -1 \\
1 & 0 & 0 & \ldots & \ldots & 0 & 0 &0 \\
0 & 1 & \ldots & 0 & \ldots & 0 & 0 &0 \\
\vdots & \vdots & \vdots & \vdots & \vdots & \vdots & \vdots & \vdots \\
0 & 0 & \ldots & \ldots & \ldots & \ldots & 0 & 0\\
0 & 0 & \ldots & \ldots & \ldots & \ldots & 1 & 0
\end{pmatrix}.
\end{align}
The transfer matrix is a $2n \times 2n$ dimensional matrix, where, we  remind that $n$ is the range of hopping. Its determinant is $1$. Interestingly, even though the system Hamiltonian is Hermitian, the transfer matrix corresponding to the lattice chain is non-Hermitian by construction \cite{universal_subdiffusive,Molinari_1997,Molinari_1998,transfer_matrix1,transfer_matrix2}, and can have EPs \cite{universal_subdiffusive,finite-range-subdiffusive,isolated_draft} of arbitrary orders. The impact of the order of the EP on quantum transport is one of the central aim of this work. 

\subsection{Eigenvalues and eigenvectors of the transfer matrix}
\label{EP_band}
In the large $N$ limit, the single particle eigenvalues and eigenvectors can be obtained via Fourier transform. The single-particle eigenvalues and the component of single particle eigenvectors are  given by 
\begin{align}
\label{dispersion}
\omega=\epsilon(k)=-2 \sum_{m=1}^n t_m \cos \big(mk\big),~~
 \psi_\ell = \frac{e^{-ik\ell}}{\sqrt{N}}, 
\end{align}
where $k$ is the lattice wave-vector and ranges between $-\pi \leq k \leq \pi$, i.e, within the Brillouin zone. It can be shown that given a value of $\omega$ satisfying $\omega=\epsilon(k)$, the transfer matrix $\mathbf{T}(\omega)$ has the following eigenvalue equation:
\begin{align}
\label{T_eigvals_eigvecs_relation}
    &\mathbf{T}(\omega)\ket{\phi^{(n)}( k)}=e^{- ik}\ket{\phi^{(n)}(k)},~~\forall~~\omega=\epsilon(k),
\end{align}
where, 
\begin{align}
\label{phi_n_k_def}
   \ket{\phi^{(n)}(k)}= \begin{pmatrix} e^{-ikn} \\ e^{-ik(n-1)}\\ \vdots \\ e^{ik(n-3)}\\ e^{ik(n-2)}\\ e^{ik(n-1)}   \end{pmatrix}.
\end{align}
Also since $\epsilon(-k) = \epsilon(k)$, 
then if $e^{- ik}$ is an eigenvalue corresponding to eigenvector $\ket{\phi^{(n)}( k)}$, then $e^{ik}$ is also an eigenvalue corresponding to the eigenvector  $\ket{\phi^{(n)}(-k)}$, i.e, 
\begin{align}
\label{T_eigvals_eigvecs_relation_minus}
    &\mathbf{T}(\omega)\ket{\phi^{(n)}( -k)}=e^{ik}\ket{\phi^{(n)}(-k)},~~\forall~~\omega=\epsilon(-k).
\end{align}
Note that, there may be only few solutions of $k$ in the Brillouin zone corresponding to a given value of $\omega$. But, the matrix $\mathbf{T}(\omega)$ has $2n$ eigenvalues. All eigenvalues of $\mathbf{T}(\omega)$ can be found by finding all complex values $z$, satisfying $\omega=\epsilon(z)$. There can be $2n$ such values of $z$. The corresponding transfer matrix eigenvalues and eigenvectors are 
\begin{equation}
   \mathbf{T}(\omega)\ket{\phi^{(n)}(z)}=e^{iz}\ket{\phi^{(n)}(z)},
\end{equation} 
thereby yielding all $2n$ eigenvalues and eigenvectors. The real values of $z$ so found, are the Bloch vectors $k$ in the Brillouin zone discussed before.


Although the above gives all the eigenvalues and eigenvectors of the transfer matrix, being non-Hermitian, the transfer matrix is not guaranteed to be diagonalizable. Parameter values where a non-Hermitian matrix is non-diagonalizable are called EPs. At an EP, several eigenvectors of the matrix coalesce. The number of eigenvectors coalescing gives the order of the EP. This is more rigourously defined via transformation to the Jordan normal form, as discussed below.  

Regardless of diagonalizability, any square matrix can be brought into a Jordan normal form via a similarity transformation, 
\begin{align}
\label{sim_trans}
    \mathbf{T}(\omega)=R J R^{-1},
\end{align}
with 
\begin{align}
\label{Jordan}
\mathbf{J}= 
\begin{bmatrix}
[\mathbf{J}]_1 & 0 &0 &\ldots &\ldots  \\
0 & [\mathbf{J}]_2 & 0 &0 &\ldots  \\
0 & 0 & [\mathbf{J}]_3 & 0 &0  \\
\vdots &\ddots &\ddots &\ddots &\vdots  \\
\ldots  &\ldots  &0 & 0 & [\mathbf{J}]_N
\end{bmatrix},
\end{align}
where $[\mathbf{J}]_i$ is a $\ell_i \times \ell_i$ square matrix,
\begin{align}
\label{def_Lamda_N}
&[\mathbf{J}]_i = \mathbf{\Lambda}^{(i)} + \mathbf{N}^{(i)},~\mathbf{\Lambda}^{(i)}= \lambda_i \mathbb{I}_{\ell_i}, \nonumber\\
& N^{(i)}_{p q} = \delta_{p,q-1}, ~~1 \leq p,q \leq \ell_i.
\end{align}
$N^{(i)}_{p q}$ are the elements of the matrix $\mathbf{N}^{(i)}$, $\mathbb{I}_{\ell_i}$ is the $\ell_i$ dimensional identity matrix. $\lambda_i$ are the eigenvalues of transfer matrix. Each eigenvalue $\lambda_i$
has a degeneracy of $\ell_i$.
The block matrices $[\mathbf{J}]_i$ are called the Jordan blocks. The size of the blocks must satisfy, $\sum_i \ell_i = 2n$, which is the size of the full transfer matrix. 
If $\mathbf{T}(\omega)$ would be diagonalizable, then $\ell_i=1$ $\forall \, i$ and $\mathbf{J}$ would be the diagonalized matrix. When $\mathbf{T}(\omega)$ is not diagonalizable, i.e, there is an EP of the transfer matrix, several eigenvalues and eigenvectors of the $\mathbf{T}(\omega)$ coalesce, leading to one or more corresponding Jordan blocks of size $\ell_i>1$. The size of each such Jordan block is equal to the number of eigenvalues and eigenvectors coalescing and gives the order of the EP.  As we will show later, this structure of expressing the transfer matrix in terms of Jordan normal form plays an instrumental role in understanding open quantum system properties. 

We next discuss engineering of such EPs of the transfer matrix for the given lattice setup in Eq.~\eqref{lattice-hamiltonian}. 
For a given range of hopping $n$, one can tune the hopping parameters $t_1, t_2, \cdots t_n$ to systematically generate EPs from $2$-nd order to $2n$-th order. 
In fact, there is a remarkable connection between transfer matrix EPs and the extrema and saddle points the energy dispersion \cite{isolated_draft}. The order of the EP, say $p$-th order, implies that the $p$-th derivative of $\epsilon(k)$ in Eq.~\eqref{dispersion} with respect to $k$ is non-zero and all lower order derivatives are zero. Utilizing this property, one can systematically generate band edges which correspond to an arbitrarily high order EP of the transfer matrix, as we discuss below.


\subsection{Band edges and transfer matrix}
\label{band_edges_and_transfer_matrix}
Band edges are global extrema of $\epsilon(k)$. Thus, they must correspond to EPs of the transfer matrix. Furthermore, being extrema the order of the EP should necessarily be even. But, there are additional conditions. The upper (lower) band edge corresponds to the global maximum (minimum). So, there should not be any Bloch wave vectors corresponding to energies infinitesimally above (below) the upper (lower) band edge. From this we have the following definition of band edge. 
\begin{definition}
	\label{band_edge_theorem1}
A given $\omega=\omega_b$ is the upper (lower) band edge if and only if, there are real values of $k$ satisfying $\omega_b=\epsilon(k)$, but, for infinitesimal $\delta \omega>0$ ($\delta \omega<0$), there are no real values of $k$ satisfying $\omega_b+\delta \omega=\epsilon(k)$. 
\end{definition}
From definition ~\ref{band_edge_theorem1} and Eq.~\eqref{T_eigvals_eigvecs_relation} , we have the following equivalent definition of band edge in terms of transfer matrix.
\begin{definition}
	\label{band_edge_theorem2}
	A given $\omega=\omega_b$ is the upper (lower) band edge if and only if, $\mathbf{T}(\omega_b)$ has some eigenvalues of unit modulus, but, $\mathbf{T}(\omega_b+\delta \omega)$, for infinitesimal $\delta \omega>0$ ($\delta\omega~<~0$), has no eigenvalue of unit modulus.   
\end{definition}
We now put forward the following important theorems.

\setcounter{lemma}{0}

\begin{theorem}
\label{band_edge_theorem3}
If $\omega=\omega_b$ is a band edge, all eigenvalues of $\mathbf{T}(\omega_b)$ that are of unit modulus are repeated even number of times, i.e, correspond to Jordan blocks of even sizes.
\end{theorem} 
{\it Proof:} Let \{$k_r\}$ be the set of real solutions for $\omega=\omega_b= \epsilon(k_r)$. We are now interested in how this solution changes for each element of the set ($k_r$) by slightly changing $\omega_b$. In other words, setting $\omega_b+\delta \omega= \epsilon(z_r)$, we want to find $z_r-k_r$ to the leading order. Taylor expanding $\epsilon(z_r)$ around $k_r$, and using $\omega_b=\epsilon(k_r)$, $\omega_b+\delta \omega= \epsilon(z_r)$, we get to the leading order in $z_r-k_r$,
\begin{align}
\label{delta_omega_a}
\delta \omega \simeq  a_p(k_r)\frac{(z_r-k_r)^p}{p!},~{\rm where},~a_p(k_r)=\frac{d^p \epsilon}{d k^p}\Big|_{k=k_r}.
\end{align} 
Here $p$ is the order of the first non-zero derivative of $\epsilon(k)$ with respect to $k$ at $k=k_r$. 
Solving for $z_r$ from Eq.~\eqref{delta_omega_a}, we get 
\begin{align}
\label{z_0}
z_r = k_r+ \left(\frac{p! \, \delta \omega}{a_p(k_r)}\right)^{1/p}.
\end{align}
Note that if $p=1$ in Eq.~\eqref{z_0}, then $z_r$ remains real on infinitesimal change of $\omega_b$. In contrast,  if $p>1$,  and $a_p(k_r)>0$ ($a_p(k_r)<0$), infinitesimally decreasing (increasing) $\omega_b$ i.e., $\delta \omega<0$ ($\delta \omega>0$) leads to $z_r$ becoming complex. 
From definition \ref{band_edge_theorem1} if $\omega_b$ is the lower (upper) band edge, this should be the case for each $z_r$. Therefore there can not be any real solutions $k_r$ that are not repeated. Moreover, by virtue of being a band edge (extrema) $p$ can never be odd. In other words, those eigenvalues of ${\bf T}(\omega_b)$ that are of unit modulus are necessarily repeated even number of times. 

Finally, we have the following theorem about the highest order EP of the transfer matrix.
\begin{theorem}
\label{highest_order_theorem}
The $2n$-th order of EP of a $2n \times 2n$ transfer matrix $\mathbf{T}(\omega)$ can occur 
only at $k=0,\pm \pi$. The corresponding value of $\omega$ must be at a band edge. 
\end{theorem}
\noindent
{\it Proof:} The 
determinant of $\mathbf{T}(\omega)$ is $1$ for any $\omega$. For $2n$-th order EP, there is only one eigenvalue, which is repeated $2n$ times. Thus, the eigenvalue has to be either $1$ or $-1$. From Eq.~\eqref{T_eigvals_eigvecs_relation}, it follows that $k=0,\pm \pi$. Furthermore, since there is a single Jordan block of size $2n$, it follows from Eq.~\eqref{z_0} that the condition for either upper or lower band edge in definition \ref{band_edge_theorem1} is satisfied. This confirms that the corresponding $\omega$ should correspond to a band edge.
\\

\noindent
We are interested in exploring how the order of the transfer matrix EP at band edge affects the nature of quantum transport. From theorem \ref{highest_order_theorem} we see that, this can be explored by considering the case with $2n$-th order transfer matrix EP and finding how the conductance scaling with system size depends on $n$. This allows for a completely general analytical treatment, as we show later. Moreover, since, for a chosen $n$, parameters required for the $2n$-th order EP can be found by taking $2n-1$ derivatives of band dispersion in Eq.~\eqref{dispersion} and equating them to zero at $k=0,\pm \pi$ \cite{isolated_draft}, exact numerical demonstration becomes also possible. Before delving into these questions, in the next section, we give the general formalism to explore the transport behavior in presence of baths.


\section{Open system setup and transport} 
\label{open}
In sec.~\eqref{isolated_transfer}, we have briefly mentioned the finite-range hopping lattice systems which host arbitrary order EP of transfer matrix and the connection between band edges and transfer matrix. In this section, we reveal how the order of the EP of the transfer matrix can show up in open system transport properties.

We consider one-dimensional lattice with a finite range of hopping $n$, as described in Sec.~\eqref{isolated_transfer} and is given by the Hamiltonian in Eq.~\eqref{lattice-hamiltonian}. The lattice system is subjected to two fermionic reservoirs at its two ends. The source bath at the left end is kept at a chemical potential $\mu_L$ and the drain bath at the right end is kept at a chemical potential $\mu_R$. The chemical potential difference between the source and the drain drives a particle current through the lattice chain. 

To further model dephasing effects, we introduce B{\"u}ttiker voltage probes (BVPs) at each lattice site. A schematic of this entire setup is presented in Fig.~\ref{fig:schematic}. These are baths similar to the source and drain baths, except that their chemical potentials, $\{\mu_n\}$, are fixed by demanding that there is no average particle current into each of them. Thus, average current still flows from source to drain.  The temperature of all the baths are considered same, given by inverse temperature $\beta$. 

The microscopic Hamiltonian $\hat{H}$ for this whole setup can be written as, 
\begin{align}
&\hat{H}=\hat{H}_C + \hat{H}_L + \hat{H}_R + \sum_{n=1}^{N} \hat{H}_{P_n}+\hat{H}_{CL} \nonumber  \\ 
& \quad \quad + \hat{H}_{CR} + \sum_{n=1}^{N} \hat{H}_{CP}^{(n)},
\end{align}
where $\hat{H}_C$ is the Hamiltonian of the lattice chain, given in Eq.~\eqref{lattice-hamiltonian}.  $\hat{H}_L$ is the Hamiltonian of the left bath (source), $\hat{H}_R$ is the Hamiltonian of the right bath (drain), $\hat{H}_{P_n}$ is the Hamiltonian of the probe attached to the $n$-th site of the system, $\hat{H}_{CL}$, $\hat{H}_{CR}$ and $\hat{H}_{CP}^{(n)}$ are the Hamiltonians which describe the coupling of central system with source, drain, and $n$-th probe respectively. Each bath is modelled via an infinite number of fermionic modes, and the system bath couplings are taken as number conserving bilinear in fermionic creation and annihilation operators, i.e, 
\begin{align}
& \hat{H}_{\ell}=\sum_{r=1}^\infty \Omega_{r \ell} \hat{B}_{r \ell}^\dagger \hat{B}_{r \ell},~~\ell=L,R,P_n, \nonumber \\
& \hat{H}_{CL}=\sum_{r=1}^\infty \left(\kappa_{r L} \hat{c}_1^\dagger \hat{B}_{r L} + \kappa_{r L}^*  \hat{B}_{r L}^\dagger \hat{c}_1 \right), \nonumber \\
& \hat{H}_{CR}=\sum_{r=1}^\infty \left(\kappa_{r R} \hat{c}_N^\dagger \hat{B}_{r R} + \kappa_{r R}^*  \hat{B}_{r R}^\dagger \hat{c}_N \right), \nonumber \\
& \hat{H}_{CP}^{(n)}=\sum_{r=1}^\infty \left(\kappa_{r P_n} \hat{c}_n^\dagger \hat{B}_{r P_n} + \kappa_{r P_n}^*  \hat{B}_{r P_n}^\dagger \hat{c}_n \right).
\end{align}
The non-equilibrium steady state (NESS) of the system in presence of all the baths can be described via the retarded non-equilibrium Green's function (NEGF) \cite{Keldysh1,Keldysh2,Keldysh3,Keldysh4,Wang2014,Wingreen}, given as
\begin{align}
\label{NEGF}
G(\omega)=\left[\omega \mathbb{I}- \mathbf{H}-\mathbf{\Sigma}_L(\omega)-\mathbf{\Sigma}_R(\omega)-\sum_{n=1}^N \mathbf{\Sigma}_{P_n}(\omega)\right]^{-1},
\end{align}
where $\mathbb{I}$ is the $N \times N$ identity matrix, $\mathbf{H}$ is the single particle Hamiltonian of the system defined in Eq.\eqref{single_particle_def}, and $\mathbf{\Sigma}_L(\omega)$,$ \mathbf{\Sigma}_R(\omega)$, $\mathbf{\Sigma}_{P_n}(\omega)$ are the retarded self-energy matrices due to the presence of the left, the right and the probe baths. The only non-zero elements in self-energy matrices are $\bra{1}\mathbf{\Sigma}_L(\omega)\ket{1}$, $ \bra{N}\mathbf{\Sigma}_R(\omega)\ket{N}$, $\bra{n}{\mathbf{\Sigma}_{P_n}(\omega)}\ket{n}$. For simplicity, we assume wide-band limit, where 
\begin{align}
\label{self_energies}
& \bra{1}\mathbf{\Sigma}_L(\omega)\ket{1}=-i\tau/2,~\bra{N}\mathbf{\Sigma}_R(\omega)\ket{N}=-i\tau/2, \nonumber \\
& \bra{n}\mathbf{\Sigma}_{P_n}(\omega)\ket{n}=-i\tau_P/2,
\end{align}
and all other elements are zero. In above, $\tau$ gives the effective coupling strength between the system and to the left and right baths, and $\tau_P$ gives the effective coupling strength between the system and to the probe baths. We have assumed coupling to all the probe baths to be same.

We take $\mu_L=\mu+\delta \mu/2$, $\mu_R=\mu-\delta \mu/2$, and zero temperature limit $\beta\to \infty$. In this setting, for small $\delta \mu$, the zero temperature conductance is given in terms of the NEGF by \cite{Pastawski,self-2-Abhishek,segal-probe-3,buttikerdephase,quasi-periodic-buttiker,superballistic1}
\begin{align}
\label{conductance}
\mathcal{G}(\mu)  = &\tau^2  \big| \bra{1}G(\mu)\ket{N} \big|^2 +\tau^2\tau_P \sum_{n,j=1}^{N}\Big[ \big|\bra{N}G (\mu)\ket{n}\big|^2 
 \nonumber \\ 
& \times \bra{n}\mathcal{W}^{-1}(\mu)\ket{j} \big| \bra{j} G (\mu) \ket{1}\big|^2\Big]. 
\end{align}
Here the elements of the  $N\times N$ matrix $\mathcal{W}(\mu)$ are
\begin{align}
\label{W}
\bra{n}\mathcal{W}(\mu)\ket{j} &= -\tau_P \big| \bra{n} G(\mu) \ket{j} \big|^2, \quad \forall \, n\neq j \nonumber \\
\bra{n}\mathcal{W}(\mu)\ket{n}  &=\tau \left(\big| \bra{n}G(\mu)\ket{1}\big|^2 +\big|\bra{n}G(\mu)\ket{N}\big|^2\right) \nonumber \\
& +\tau_P  \sum_{j\neq n}^{N}\big|\bra{n}G(\mu)\ket{j}\big|^2.
\end{align}
It is important to note that Eqs.~\eqref{conductance} and \eqref{W} are exact and non-perturbative in nature.  How conductance scales with system size $N$ defines the nature of transport in open systems. For normal diffusive transport, $\mathcal{G}(\mu)\sim N^{-1}$, such that, conductivity, given as $\sigma(\mu)=\lim_{N\to\infty}N \mathcal{G}(\mu)$ is well-defined. For ballistic transport, conductance does not scale with $N$, $\mathcal{G}(\mu)\sim N^0$. If  $\mathcal{G}(\mu)\sim N^{-\delta}$, with $\delta\neq 0,1$, we have anomalous transport. If $0<\delta<1$, transport is called superdiffusive, while if $\delta>1$, transport is called subdiffusive. Finally, if there is complete lack of transport, for example in Anderson localized case, conductance decays exponentially with system size, $\mathcal{G}(\mu)\sim e^{-N/\xi}$, where $\xi$ is the localization length. It is noteworthy that if $\delta <0$, the the transport can be dubbed as superballistic which was recently reported \cite{superballistic1}. Apart from this fundamental importance, knowing the behavior of conductance with system length also has technological implications in nanoscale devices, where such finite-size behavior can become important. Our central goal here is to reveal how the higher order EPs of transfer matrices affect the scaling of conductance with system size, both in absence and in presence of bulk dephasing. 

Previously, in Ref.~\cite{universal_subdiffusive}, we considered a nearest-neighbour hopping system and showed the connection between conductance scaling with system length and second-order EPs of the transfer matrix at band edges, in absence of the BVPs, i.e, $\tau_P=0$. We denote by $\mathcal{G}^{(0)}(\mu) $ the conductance when $\tau_P=0$. We found that, conductance shows a universal subdiffusive scaling,  $\mathcal{G}^{(0)}(\mu) \sim N^{-2}$ at the band edges. In Ref.~\cite{finite-range-subdiffusive}, some of the authors studied a particular finite-range hopping case which had a fourth order EP at a band edge and still observed the same $\mathcal{G}^{(0)}(\mu) \sim N^{-2}$ behavior. A natural question to ask how such $N^{-2}$ subdiffusive scaling of conductance is affected by the order of the EP of the transfer matrix.  In the next section, we investigate this question. 
It is worth noting that such $N^{-2}$ scaling at the band edges has also been reported even for systems with no straightforward connection to EPs (quasi-1D system (extended ladder) \cite{junaid_subdiffusive} and long-range hopping systems \cite{subdiffusive_long_range}).

\section{Boundary driven transport in absence of bulk dephasing}
\label{probe_absence}
This section is devoted for studying transport through the lattice system presented in Fig.~\ref{fig:schematic} but in the absence of bulk dephasing probes i.e, $\tau_p$ is set to zero. 

\subsection{Relation between conductance scaling and transfer matrix}
The conductance in absence of bulk dephasing [setting $\tau_p=0$ in Eq.~\eqref{conductance}] is given by 
\begin{align}
 \mathcal{G}^{(0)}(\mu)  = \tau^2  \big| \bra{1} G^{(0)}(\mu) \ket{N} \big|^2, 
 \label{cond}
\end{align}
where $G^{(0)}(\mu)$ is the NEGF with $\tau_P=0$ [see Eqs.~\eqref{NEGF}, \eqref{self_energies}]. Since the baths are only attached at the first and the $N$-th sites, the $N$ dependence of retarded NEGF is expected to be the same as that of $\bra{1} g(\mu) \ket{N}$, where
\begin{align}
\label{bare_greens_function}
g(\mu)=\lim_{\eta\to 0}\big[(\mu+i\eta)\mathbb{I}-\mathbf{H} \big]^{-1},
\end{align} 
is the bare retarded Green's function in absence of the baths. So we have,
\begin{align}
\mathcal{G}^{0}(\mu)\propto \big|\bra{1} G^{(0)}(\mu) \ket{N} \big|^2 \propto \big|\bra{1} g(\mu)\ket{N} \big|^2, \label{coherent_bare}
\end{align}
in absence of any BVPs. Noting that $\mathbf{H}$ is a Toeplitz matrix for a finite-range hopping system, the relation between the matrix elements of the bare Green's function, $\bra{i} g(\mu) \ket{j}$ and the transfer matrix $\mathbf{T}(\mu)$ is given via the following equations \cite{finite-range-subdiffusive},
\begin{equation}
\bra{i}g(\mu)\ket{j}=\frac{\bra{i}\big[\mathbf{M}(\mu)^{-1}\big]\ket{j}}{t_n}, \label{g_M}
\end{equation}
where, 
\begin{widetext}
\begin{align}
&\langle i|\textbf{M}(\mu)^{-1}|j \rangle= 
\begin{cases}
   \sum\limits_{m=1}^n \left\langle n \vert \mathbf{T}(\mu)^{-i} \vert n+m \right \rangle \langle m|\textbf{M}(\mu)^{-1}|j \rangle,& 
   \text{if }~ j>i\\
    \sum\limits_{m=1}^n \left\langle n \vert \mathbf{T}(\mu)^{-i} \vert n+m \right \rangle \langle m|\textbf{M}(\mu)^{-1}|j\rangle- \left\langle n \vert \mathbf{T}(\mu)^{-(i-j+1)} \vert 1 \right\rangle,  & \text{if}~ j\leq i 
\end{cases},\label{M1} 
\\
\label{M2a}
&  \sum\limits_{m=1}^n \bra{s}\mathbf{A}(\mu)\ket{m} \, \langle m|\textbf{M}(\mu)^{-1}|j \rangle 
 -\langle s+n \vert \mathbf{T}(\mu)^{-(N-j+1)} \vert 1 \rangle =0, \quad {\rm {where}}  \\
 & \quad \quad \bra{s}\mathbf{A}(\mu)\ket{m}=\langle s+n \vert \mathbf{T}(\mu)^{-N} \vert n+m \rangle, ~~m,s=1,2,\ldots,n.\label{M2b}
\end{align}
\end{widetext}
We emphasis that the procedure to calculate matrix elements of $\textbf{M}(\mu)^{-1}$ is as follows. Given $\mathbf{T}(\mu)$ which is a $2n \times 2n$ matrix (recall that $n$ is the range of hopping), we use Eq.~\eqref{M2b} to get matrix elements of $\mathbf{A}(\mu)$ which is a $n \times n$ matrix. We then use Eq.~\eqref{M2a} to obtain the matrix elements $\bra{m} \mathbf{M}(\mu)^{-1} \ket{j}$ for $m=1,2, \cdots n$ and for any $j$. The remaining matrix elements of $\mathbf{M}(\mu)^{-1}$ can then be obtained following Eq.~\eqref{M1}.
Following Eq.~\eqref{coherent_bare}, to calculate the conductance, we only need the $\bra{1} g(\mu)\ket{N}$, which after some algebraic manipulation further simplifies to
\begin{align}
\label{g1na2a}
\bra{1}g(\mu)\ket{N}=-\frac{1}{t_n} \left \langle 1 \vert \mathbf{A}^{-1}(\mu) \vert n \right \rangle.
\end{align}
Equations \eqref{g_M}, \eqref{M1}, \eqref{M2a}, \eqref{M2b}, along with Eq.~\eqref{g1na2a} give the connection between the conductance scaling with system length and  and the transfer matrix of the lattice, in absence of BVPs.
Remarkably, albeit ${\bf A(\mu)}$ is a $n\times n$ matrix, its relation in terms of  transfer matrix ${\bf T}(\mu)$ is given in Eq.~\eqref{M2b} which explicitly has system size dependence encoded in it. This is what leads to highly non-trivial conductance scaling with system size which is discussed in the following subsections. 
\begin{figure*}
\includegraphics[width=\textwidth]{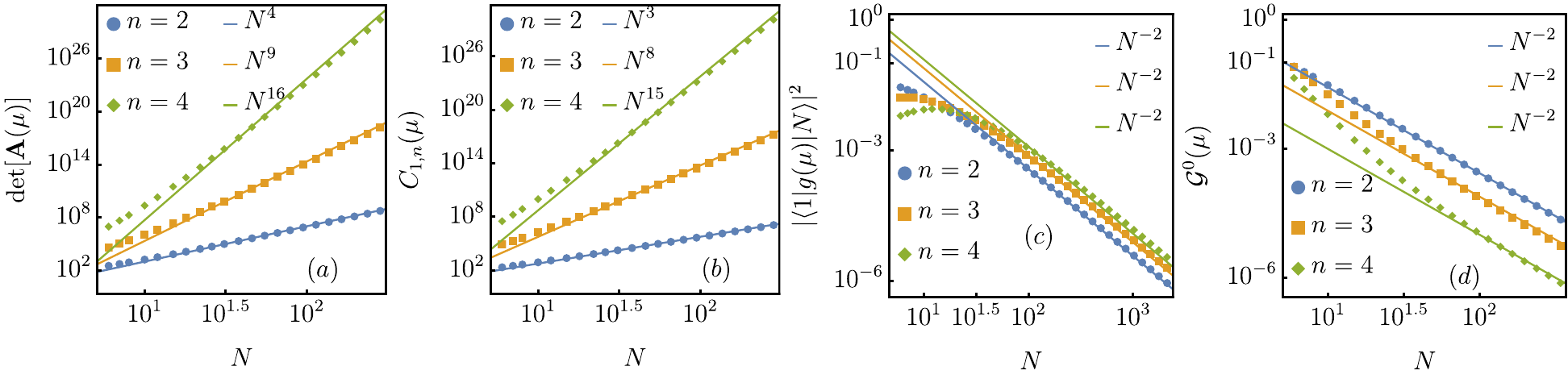}
\caption{(a) The figure shows scaling of determinant of $\mathbf{A(\mu)}$, given in Eq.~\eqref{M2a} and Eq.~\eqref{M2b} with $N$, when the $2n \times 2n$ transfer matrix $\mathbf{T}(\mu)$ has a $2n-$th order EP. (b) The figure shows the scaling of the co-factor of the $1,n$ element of $\mathbf{A}(\mu)$, denoted by $C_{1,n}(\mu)$ [see Eq.~\eqref{A_inv_C}], with $N$, when the $2n \times 2n$ transfer matrix $\mathbf{T}(\mu)$ has a $2n$th order EP. (c) The figure shows plots of $\big|\bra{1} g(\mu) \ket{N}\big|^2=\big|\frac{1}{t_n} \left \langle 1 \vert \mathbf{A}^{-1}(\mu) \vert n \right \rangle\big|^2$
with $N$, when there is a $2n-$th  order EP of $\mathbf{T}(\mu)$. Irrespective of the range of hopping $n$, the scaling with system size is $1/N^2$. (d) The figure shows plots of scaling of  $\mathcal{G}^{(0)}(\mu)$, defined in Eq.~\eqref{cond}, with $N$ when $\mu$ and the hopping strengths are such that there is a $2n-$th   order EP of the transfer matrix. Regardless of the value of $n$, we find $\mathcal{G}^{(0)}(\mu) \sim N^{-2}$. For $n=2$, $t_2=1/4$; for $n=3$, $t_2=2/5$, $t_3=1/15$; for $n=4$, $t_2=1/2$, $t_3=1/7$, $t_4=1/56$. For all cases, $\mu=-2 \sum_{m=1}^n (-1)^m t_m$.   
\label{fig:subdiffusive}}   
\end{figure*}

\subsection{Conductance scaling away from transfer matrix EPs}
We first discuss the scenario where $\mu$ is such that the transfer matrix $\mathbf{T}(\mu) $ is away from any EP i.e., $\mu$ does not correspond to the band extrema. 
$\mathbf{T}(\mu)^{-N}$ in Eq.~\eqref{M2a} and Eq.~\eqref{M2b} can then be calculated by diagonalizing the transfer matrix $\mathbf{T}(\mu)$. As discussed before, eigenvalues of transfer matrix come in pairs of the form $e^{\pm iz}$. If some values of $z$ satisfying $\mu=\epsilon(z)$ are real, which can happen when the chemical potential is within the system band, then, $z=k$, with $k$ being the Bloch wave momentum. In this case, $\mathbf{T}(\mu)^{-N}$ does not scale with $N$ but rather oscillates with $N$. Consequently, following Eqs. \eqref{coherent_bare} and \eqref{g1na2a}, the conductance also oscillates with $N$, without showing any system size scaling behavior. This is the hallmark of ballistic transport, where, $\mathcal{G}^{(0)}(\mu) \sim N^0$.     

If all values of $z$ satisfying $\mu=\epsilon(z)$ are complex, then the chemical potential $\mu$ lies outside the system band. In this case, the transfer matrix can once again be diagonalized and the norm of the matrix $\mathbf{T}(\mu)^{-N}$ diverges exponentially. Consequently, every element of $\mathbf{A}(\mu)$ diverges exponentially, and hence, those for $\mathbf{A}^{-1}(\mu)$ decay exponentially. Following Eqs.~\eqref{coherent_bare} and \eqref{g1na2a}, this shows that the conductance decays exponentially, $\mathcal{G}^{(0)}(\mu) \sim e^{-N/\xi}$ where $\xi$ is the localization length. This behavior happens when chemical potential $\mu$ is outside the system bands.

The above discussion tells us that any scaling of conductance with system size different from the above two behaviors (within and outside the band) requires $\mu$ to be such that $\mathbf{T}(\mu) $ is not diagonalizable, and hence at an EP. One may then expect that there will be different types of conductance scaling with system size, depending on the    order of the EP. Despite this, as we will show below, it remarkably turns out that for transport,in absence of the BVPs, the only other behavior of conductance is $\mathcal{G}^{(0)}(\mu) \sim N^{-2}$ when $\mu$ corresponds to band edges, and  is insensitive to the order of the EP.

\subsection{Conductance scaling in the presence of highest order EP of transfer matrix}
\label{sub:immunity}
We first consider the case where the transfer matrix has the highest order EP. For the given setup, the highest order EP can only occur when $\mu$ is located at a band edge and $k=0$ or $k=\pm \pi$ \cite{isolated_draft}. We assume $k=\pi$ in the following, without loss of generality. Recall that, for our system with the finite range of hopping $n$, the transfer matrix is $2n \times 2n$. So, the highest order EP is of order $2n$.  As explained in Sec.~\ref{EP_band}, at the EPs the transfer matrix is not diagonalizable but can be similarity transformed into a Jordan normal form $J$, $\mathbf{T}(\mu)=R J R^{-1}$, so that, we have
\begin{align}
\mathbf{T}(\mu)^{-N}=R \, J^{-N} \, R^{-1}.
\end{align}
Given that $\mathbf{T}(\mu)$ has a $2n$-th    order EP at the chosen $\mu= \epsilon(k=\pi)$, the expression for the elements of $\mathbf{T}(\mu)^{-N}$ are obtained as,
\begin{align}
\label{Tmata}
 \left\langle k \vert \mathbf{T}^{-N}(\mu) \vert m \right\rangle=\sum\limits_{s=1}^{2n} \sum\limits_{i=1}^{2n+1-s}  \left\langle 1 \vert J^{-N} \vert s \right\rangle \left\langle k \vert R \vert i \right\rangle \\ \nonumber
  \times \left\langle i+s-1 \vert R^{-1} \vert m \right\rangle,
\end{align}
where
\begin{align}
\left \langle 1 \vert J^{-N} \vert 1 \right \rangle&=(-1)^{-N}, \nonumber \\ 
\left \langle 1 \vert J^{-N} \vert s \right \rangle&=(-1)^{-N} \prod\limits_{i=1}^{s-1} \frac{N+s-i-1}{i} ~\forall~ s=2~ \mathrm{to}~ 2n
\label{Jele}
\end{align}
and elements of $2n \times 2n$ matrix $R$ are given by
\begin{align}
\label{R}
& \bra{s} R \ket{1}=(-1)^s,~~\forall~~s=1~{\rm to}~2n \nonumber \\
& \bra{2n} R \ket{j}=0,~~~~~~\forall~~j=s~{\rm to}~2n \nonumber \\
& \bra{s} R \ket{j}\!=\!(-1)^{s+j-1}\Big( \big| \bra{i\!+1} R \ket{j} \big| \!+\! \big| \bra{i\!+1} R \ket{j\!-1} \big| \Big), \nonumber \\
& \quad \quad \quad \qquad \qquad ~~~~~ \forall~~s=1~{\rm to}~2n-1 ~~{\rm and} ~~j=2~{\rm to}~2n. \nonumber \\
\end{align}
and those of $R^{-1}$ are given by
\begin{align}
\label{R_inv}
& \bra{s} R^{-1} \ket{2n}=1,~~\forall~~s=1~{\rm to}~2n\nonumber \\
& \bra{1} R^{-1} \ket{j}=0,~~~~\forall~~j=1~{\rm to}~2n-1 \nonumber \\
& \bra{s} R^{-1} \ket{j}= \bra{i-1} R^{-1} \ket{j} + \bra{i-1} R^{-1} \ket{j+1} \nonumber \\
&\qquad \qquad \qquad ~~~~~~~\forall~~j=1~{\rm to}~2n-1~~{\rm and}~~s=2~{\rm to}~2n. \nonumber \\
\end{align}
Interestingly, $R$ and $R^{-1}$ matrices in Eq.~\eqref{R} and Eq.~\eqref{R_inv} have forms reminiscent of Pascal's triangle (or {\it Pingala's meru-prastara}).

To find the conductance scaling with system size, from Eqs.~\eqref{coherent_bare} and \eqref{g1na2a}, we need the $1,n-$th element of $\mathbf{A}^{-1}(\mu)$. This is given by,
\begin{align}
\left \langle 1 \vert \mathbf{A}^{-1}(\mu) \vert n \right \rangle= \frac{C_{1,n}(\mu)}{{\rm det}[\mathbf{A}(\mu)]},
\label{A_inv_C}
\end{align}
where $C_{1,n}(\mu)$ is the co-factor of the $1,n$th element of $\mathbf{A}(\mu)$, and ${\rm det}[\mathbf{A}(\mu)]$ is the determinant of the matrix $\mathbf{A}(\mu)$. From Eqs.~\eqref{M2a}, \eqref{M2b}, \eqref{Jele}, \eqref{R}, and  \eqref{R_inv}, we can find all elements of $\mathbf{A}(\mu)$ and hence find $C_{1,n}(\mu)$ and ${\rm det}[\mathbf{A}(\mu)]$. Using this, we numerically check the scaling of $C_{1,n}(\mu)$ and ${\rm det}[\mathbf{A}(\mu)]$ with $N$ for different range of hoppings $n$, and find that
\begin{align}
C_{1,n}(\mu) \sim N^{n^2-1},~~{\rm det}[\mathbf{A}(\mu)]\sim N^{n^2},
\label{scaling}
\end{align}
which is shown in Fig.~\eqref{fig:subdiffusive}(a) and (b). Using Eq.~\eqref{scaling} in Eq.~\eqref{A_inv_C}, we can see that $\langle 1|A^{-1}(\mu)|n\rangle \sim 1/N$ and remarkably is independent of the range of hopping $n$. Thus, irrespective of $n$, the conductance in Eq.~\eqref{coherent_bare} at the band edge is 
\begin{equation} 
\mathcal{G}^{(0)}(\mu) \sim|\bra{1}g(\mu)\ket{N}|^2 \sim \frac{1}{N^{2}},
\end{equation}
which is clearly demonstrated in Fig.~\eqref{fig:subdiffusive}(c). 
It is therefore clear that $\mathcal{G}^{(0)}(\mu) \sim N^{-2}$ irrespective of the  value of the highest order of EP (i.e, $2n-$th) of the transfer matrix. Note that for this analysis, we did not require to know any explicit parameters of the lattice and only relied on the fact that the transfer matrix has the highest order EP which always occurs at the band edge. This shows the generality of the result. In Fig.~\eqref{fig:subdiffusive}(d), we show the scaling of $\mathcal{G}^{(0)}(\mu) $ with $N$ via direct numerical calculation of $\tau^2  \big| \bra{1}G(\mu)\ket{N} \big|^2$ [see Eq.\eqref{conductance}], without relying on above analytical results, for some explicit cases with $n=2,3, 4$, choosing hopping parameters and $\mu$ that lead to $2n-$th order EP of transfer matrix.

\subsection{Conductance scaling in the presence of any arbitrary order of EP}

As mentioned in Sec.~\eqref{EP_band}, the $2n \times 2n$ matrix $\mathbf{T}(\mu)$, can have any order of EP, starting from second order to $2n-$th order. For EP of any order $q$ less than $2n$, the Jordan normal form is block diagonal, with non-diagonal blocks of size $q$ corresponding to the degenerate eigenvalues, the remaining part of the matrix being diagonal with diagonal elements corresponding to the non-degenerate eigenvalues. Calculating conductance requires evaluating $\mathbf{T}(\mu)^{-N}$, for which we can consider the scaling of each Jordan block.

Let us first consider a situation where $\mu=\epsilon(k)$ can be satisfied for multiple real values of $k$. Furthermore let us assume that amongst these $k$ values, some of them correspond to EP of the transfer matrix with order $q$ (recall $q$ is less than $2n$ here). This scenario can occur when the chemical potential $\mu$ lies within the system energy band.  
In this case, some of the non-degenerate eigenvalues of the transfer matrix $\mathbf{T}(\mu)$ are in complex conjugate pairs with absolute value $1$. Then the contribution from such eigenvalues will be oscillatory, leading to no scaling behavior with $N$. In this case, the overall value of the matrix element $\bra{1} g(\mu) \ket{N}$ also do not scale with $N$, leading to ballistic behavior for the conductance i.e, $\mathcal{G}^{(0)}(\mu) \sim N^{0}$. 
The contribution from the degenerate blocks occurring due to the EP, and the contribution from non-degenerate eigenvalues with absolute value different from $1$ are completely suppressed leading to overall ballistic $N^{0}$ scaling of conductance. 

Now, let us consider the case when $\mu$ is located  at a band edge. In such a scenario every eigenvalue of $\mathbf{T}(\mu)$ with absolute value $1$ is degenerate and repeated an even number of times. This follows from  Sec.~\ref{band_edges_and_transfer_matrix}, Theorem \ref{band_edge_theorem3}.  Also, all the other eigenvalues of the transfer matrix have absolute values different from $1$. In this case, in calculating the matrix element related to the conductance i.e, $\bra{1} g(\mu) \ket{N}$, the contribution from all non-degenerate eigenvalues  decay exponentially, and the sole  contribution comes from the non-diagonal Jordan blocks. This yields  $\mathcal{G}^{(0)}(\mu) \sim N^{-2}$.

Finally, when $\mu$ is located outside the band, solving $\mu=\epsilon(k)$ yields only complex solutions of $k$. As a result, all eigenvalues of the transfer matrix have absolute value different from $1$. This consequently results in $\mathcal{G}^{(0)}(\mu) $ decaying exponentially decaying with system size.


The above discussion reveals that, in absence of bulk dephasing, the scaling of conductance $\mathcal{G}^{(0)}(\mu) $ with system size $N$ in presence of any order of transfer matrix EP, is completely immune to the order of the EP. This is true irrespective of whether the chemical potential $\mu$ lies within, at, or outside the energy band of the system. In the next section, we unravel see the sensitivity of conductance $\mathcal{G}^{(0)}(\mu) $ on the   order of the transfer matrix EP in presence of bulk dephasing by introducing BVPs.

\section{Boundary driven transport in presence of bulk dephasing: Impact of the order of transfer matrix EP}
\label{with_probes}
\begin{figure*}
\includegraphics[width=2\columnwidth]{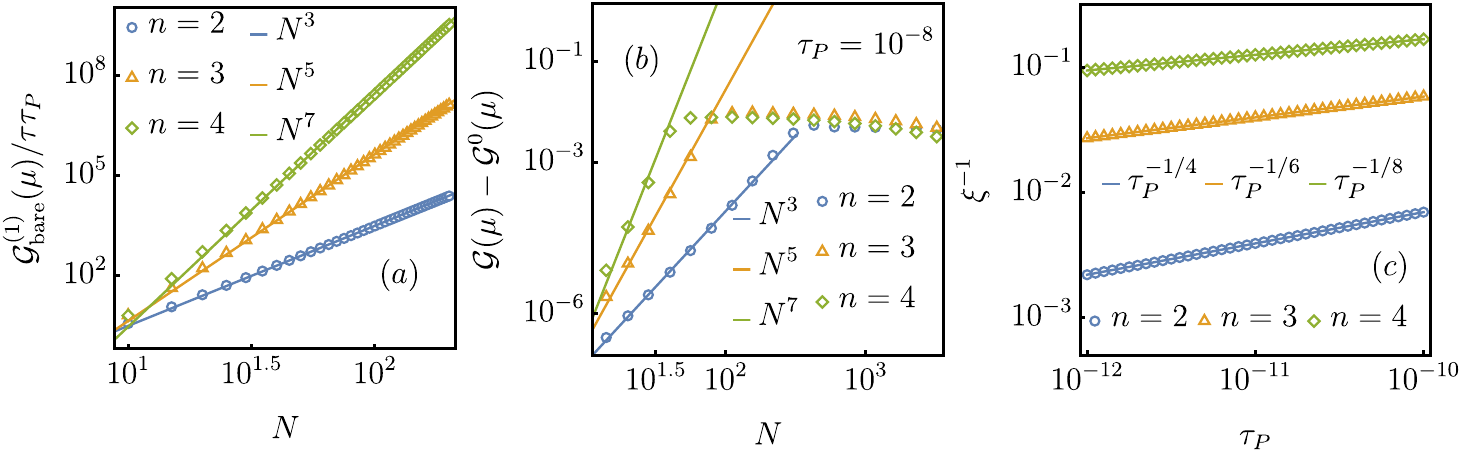} 
\caption{(a) Plot for $\mathcal{G}_{\rm bare}^{(1)}(\mu)$ in Eq.~\eqref{term1} in presence of BVPs. The conductance shows the system size scaling as $N^{2n-1}$. (b) The exact $\mathcal{G}(\mu)$ in Eq.~\eqref{conductance} with $\mathcal{G}^{(0)}(\mu)$ subtracted plotted as a function of system size $N$.
We can see that $\mathcal{G}(\mu)-\mathcal{G}^{(0)}(\mu) $ shows the superballistic scaling of conductance up to some finite size $N_{SB}^{(2)}$ with system size scaling $N^{2n-1}$. (c) Plot for the phase coherent length $\xi^{-1}\sim |\log[\lambda_{+}]|\sim \tau_P^{1/2n}$ as a function of $\tau_P$.  Here, the    order of transfer matrix EP is $2n$.}
\label{fig3} 
\end{figure*}

In this section, we discuss the consequence of order of the transfer matrix exceptional points on transport in the presence of BVPs. We recall here that
in Ref.~\cite{superballistic1}, we considered the nearest neighbour hopping case $(n=1)$, in the presence of the BVPs.  Most remarkably,  we found the counter-intuitive superballistic scaling of conductance at the band edges. Note that the term superballistic transport refers to an increase in conductance with system length as a power law, over an extended but finite regime of system sizes. This behaviour was found to be quite robust against small deviations from band edges, weak disorder, and the presence of a small but finite temperature of the boundary reservoirs and the BVPs. The superballistic scaling of conductance at the band edges that correspond to second order EP was enabled by weak coupling to the BVPs and did not occur in their absence. In the following, we study the effect of higher order EPs of transfer matrix on the superballistic scaling, and reveal that the exponent of the power-law growth with system size is remarkably governed by the order of the EP.

\subsection{Relation between conductance and EP of transfer matrix in presence of bulk dephasing}

In this subsection, we would like to analyze the conductance scaling, in the presence of BVPs, at the band edges that correspond to the higher    order EPs of the underlying transfer matrix. A promising route to tackle this problem is to adopt a perturbative treatment in the system-probe coupling strength i.e., $\tau_P$. The conductance in presence of BVPs, given in Eq.~\eqref{conductance}, in the small $\tau_P$ limit reduces to,
\begin{align}
\label{conductance2}
&\mathcal{G}(\mu) = \mathcal{G}^{(0)}(\mu) + \mathcal{G}^{(1)}(\mu) + \mathcal{O}(\tau_P^2), 
\end{align}
where 
\begin{align}
\label{conductance3}
&\mathcal{G}^{(1)}(\mu)=\tau \tau_P \sum_{\ell=1}^{N} \! \frac{|\langle \ell | G^{(0)}(\mu)| N \rangle|^2 \,  |\langle 1|G^{(0)}(\mu)|\ell \rangle|^2}{|\langle \ell | G^{(0)}(\mu)| N \rangle|^2 + |\langle 1|G^{(0)}(\mu)|\ell\rangle|^2}.
\end{align}
In above, $\mathcal{G}^{(1)}(\mu)$ is the term proportional to $\tau_P$
and $G^{(0)}(\mu)$ is the NEGF with $\tau_P=0$ [see Eqs~\eqref{NEGF},\eqref{self_energies}]. Recall that ${\cal G}^{0}(\mu)$ shows universal subdiffusive $1/N^2$ scaling and is independent of the order of transfer matrix EP, as discussed in the preceding subsection. The term $\mathcal{G}^{(1)}(\mu)$ in Eq.~\eqref{conductance3} is the key to finding the onset of superballistic scaling of conductance. For the nearest neighbor hopping model, it was earlier shown that this particular term scales as $N$ at the band edges \cite{superballistic1} and can be explained in terms of second order EP of the transfer matrix.

In what follows, we will calculate this $\mathcal{G}^{(1)}(\mu)$ in presence of higher    order
EP of the transfer matrix and discuss the impact on superballistic transport. Note that as the system size scaling of conductance is not expected to depend on the self-energies of the baths,  we further simplify $\mathcal{G}^{(1)}(\mu)$ by merely using the bare Greens functions of the system. In other words, the system size scaling, encoded in Eq.~\eqref{conductance3}, can be understood by studying the system size scaling of the following term
\begin{align}
\label{term1}
\mathcal{G}^{(1)}_{\rm bare}(\mu)=\tau \tau_P \sum_{\ell=1}^{N} \! \frac{|\langle \ell | g(\mu)| N \rangle|^2  \, |\langle 1|g(\mu)|\ell \rangle|^2}{|\langle \ell | g(\mu)| N \rangle|^2 + |\langle 1|g(\mu)|\ell \rangle|^2},
\end{align} 
where $g(\mu)$ is the bare Green's function of the system, defined in Eq.~\eqref{bare_greens_function}.
We  write down the expressions for $\langle \ell | g(\mu)| N \rangle$ and $\langle 1 | g(\mu)| \ell \rangle$ for finite-range hopping model with range $n$ using Eq.~\eqref{g_M}, Eq.~\eqref{M1}, ~\eqref{M2a}, and \eqref{M2b} as,
\begin{align}
\label{glng1l}
&\langle \ell | g(\mu)| N \rangle =-\frac{1}{t_n} \sum\limits_{m=1}^n \left\langle n \vert \mathbf{T}^{-\ell} \vert n+m \right \rangle \left\langle m|\textbf{A}^{-1}|n\right \rangle, \nonumber \\
& \langle 1 | g(\mu)| \ell \rangle=\frac{1}{t_n}  \left\langle 1 \vert \mathbf{A}^{-1} \mathbf{B}|1\right \rangle,
\end{align}
where, 
\begin{align}
\label{B-vector}
&\mathbf{B}^T =\begin{pmatrix}
\left\langle n+1 \vert \mathbf{T}^{-(N-\ell+1)} \vert 1  \right \rangle & \ldots & \left\langle 2n \vert \mathbf{T}^{-(N-\ell+1)} \vert 1  \right \rangle
\end{pmatrix},
\end{align}
where we have suppressed the argument $\mu$ of the transfer matrix for notational brevity.
As the matrix $\bf{A}$ and $\bf{B}$ can be written in terms of transfer matrix elements, all the Green's function can be expressed in terms of transfer matrix elements. Thus, the above expression clearly shows the relation between scaling of  $\mathcal{G}_{\rm bare}^{(1)}(\mu)$ and transfer matrix $\bf{T}(\mu)$.

\subsection{Superballistic scaling in $\mathcal{G}^{(1)}_{\rm bare}(\mu)$ in presence of highest order EP of transfer matrix}
Considering the transfer matrix of dimension $2n \times 2n$ has a $2n-$th    order EP, we can find the elements of transfer matrix and the Green's function using Eq.~\eqref{Tmata} and Eq.~\eqref{glng1l}. Knowing this, we can easily calculate $\mathcal{G}^{(1)}_{\rm bare}(\mu)$ in Eq.~\eqref{term1}. In fact, in Fig.~\eqref{fig3}(a), we have plotted $\mathcal{G}^{(1)}_{\rm bare}(\mu)$ considering the highest order EP of transfer matrix and interestingly find that it depends on the order of EP as 
\begin{equation}
\mathcal{G}^{(1)}_{\rm bare}(\mu) \sim N^{2n-1}.
\label{super-scaling}
\end{equation} 
In Appendix. \eqref{AppendixA}, we provide a detailed analytical derivation for the conductance scaling at the highest order EP that occurs at the band edge, both in absence [see Appendix.~\eqref{immunity_fourth}] and in presence of the BVPs [see Appendix.~\eqref{lack_immunity}].


\subsection{Window of superballistic scaling of conductance in presence of highest order EP}
If we see the behavior of $\mathcal{G}(\mu)$ in Eq.~\eqref{conductance2} in the presence of probes in the small $\tau_P$ limit at the EP, that corresponds to the band edges, the first term in the conductance shows a subdiffusive behavior with $1/N^2$ scaling as $\mathcal{G}^{(0)}(\mu) $ gives the major contribution in $\mathcal{G}(\mu)$. Also, recall from Sec.~\ref{sub:immunity} that $\mathcal{G}^{(0)}(\mu) $ is immune to the order of the EP and always shows $1/N^2$ scaling. For a fixed $\tau_P$, with increasing system size, when $\mathcal{G}^{(1)}(\mu)>\mathcal{G}^{(0)}(\mu) $, then we see the onset of the superballistic scaling. Let us denote the corresponding system size as $N_{SB}^{(1)}$. In presence of $2n-$th order EP of transfer matrix, $\mathcal{G}^{(1)}(\mu) \sim N^{2n-1}$ [see Eq.~\eqref{super-scaling} and Fig.~\eqref{fig3} (a)]. From Eq.~\eqref{conductance2}, by comparing the terms $\mathcal{G}^{(0)}(\mu)$ and $\mathcal{G}^{(1)}(\mu)$, $N_{SB}^{(1)}$ can be estimated to depend on $\tau_P$ as 
\begin{equation}
N_{SB}^{(1)}\sim \Big(\frac{1}{\tau_P}\Big)^{\frac{1}{(2n+1)}}.
\label{SB-starting}
\end{equation} 
On the other hand, let us consider the exact quantity $\mathcal{G}(\mu)$ as given in Eq.~\eqref{conductance}. When this quantity is studied with $\mathcal{G}^{(0)}(\mu) $ subtracted, i.e, $\mathcal{G}(\mu)-\mathcal{G}^{(0)}(\mu)$ we find that, starting from the smallest system sizes up to some system size, say, $N_{SB}^{(2)}$, we observe a major effect of $\mathcal{G}^{(1)}(\mu)$ which shows system size scaling as $N^{2n-1}$. This is demonstrated in  Fig.~\eqref{fig3}(b). 
For the finite range hopping model, in the presence of probes, in the large system size limit, all the Green's functions follow an exponential decay such as $|\langle i | G| j \rangle |^2 \sim e^{-|i-j|/\xi}$, where $\xi$ is the phase-coherence length \cite{superballistic1}. For system lengths much larger than $\xi$, it can be shown that conductance follows a diffusive scaling. Following Ref.~\onlinecite{superballistic1}, $N_{SB}^{(2)}\sim \xi$. The dependence of phase-coherence length on the dephasing strength $\tau_P$ then relates $N_{SB}^{(2)}$ and $\tau_P$. 

The phase-coherence length is related to the highest magnitude eigenvalue of the transfer matrix, $\lambda_{+}$  as, $\xi^{-1} \propto |\log[\lambda_{+}]|$. In presence of $2n-$th order EP of transfer matrix, $|\log[\lambda_{+}]| \sim \tau_P^{1/2n}$ [see Appendix.~\eqref{largest_eigenvalue}]. We have shown this in Fig.~\eqref{fig3}(c). Thus, 
\begin{equation}
    N_{SB}^{(2)} \propto \xi \sim \tau_P^{-1/2n}.
\end{equation} 
So, in presence of $2n-$th  order EP of transfer matrix, the window of superballistic scaling of conductance shows the $\tau_P$ dependence as $N_{SB}^{(2)}-N_{SB}^{(1)}\sim \tau_P^{-1/2n}$. Thus, for a given $\tau_P$, with increasing the order of transfer matrix EP, the value of the exponent of superballistic scaling increases but the window of superballistic scaling of conductance decreases. 
\\

\begin{figure}
\includegraphics[width=\columnwidth]{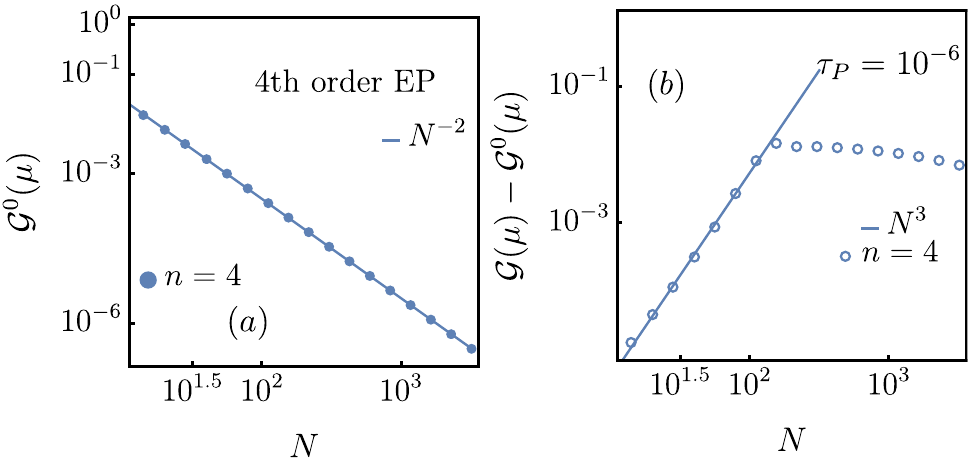} 
\caption{Fermionic transport behaviour in (a) absence and (b) presence of BVPs at the band edge when the transfer matrix has EP of order $p$, where $2<p<2n$. In above, we have chosen $n=4$ and EP is of order $p=4$, which thus satisfies $2<p<8$.}
\label{app:fig1a} 
\end{figure}

\subsection{Superballistic transport when the order of EP of the transfer matrix is less than $2n$}
We have discussed the highest possible order of EP of the transfer matrix which is $2n$ for a finite-range hopping model with range $n$. This always occurs at the band edges of the system. We now discuss one example where the order of transfer matrix EP is less than $2n$ and larger than $2$. Such cases can also occur at the band edge of the system. 
Let us consider the situation with $n=4$. In this case, the highest possible order of the transfer matrix EP is $8$ and the lowest possible order of EP is $2$. Using the recent results in Ref.~\onlinecite{isolated_draft}, we find the condition for generating a fourth-order EP of transfer matrix $\mathbf{T}(\epsilon(\pi))$ with $k=\pi$ and $\omega=\epsilon(\pi)$ as,
$t_2=(1+9t_3-16 t_4)/4$.  
Thus, this example shows the exsistance of fourth-order exceptional hypersurface of transfer matrix which lies on $(\omega,t_3,t_4)$ parameter space. Choosing a sample point from this hypersurface with $t_3=1/6$ and $t_4=1/16$, we find that, in absence of the BVPs the conductance $\mathcal{G}^0 (\mu)$ in Eq.~\eqref{cond} scales as $1/N^2$ [Fig.~(\ref{app:fig1a}a)].
In the presence of BVPs, upto a finite system size $N$, $\mathcal{G}(\mu)-\mathcal{G}^0(\mu)$ shows the remarkable $N^3$ scaling [Fig.~(\ref{app:fig1a}b)].

In general, if the band edges correspond to the $p-$th    order EP of transfer matrix, then we will always see the superballistic scaling of conductance in presence of BVPs 
\begin{equation}
\label{eq:sbp}
\mathcal{G}(\mu)-\mathcal{G}^0(\mu) \sim N^{p-1}\, ,
\end{equation}
with 
\begin{equation}
\label{eq:nsb12}
N_{SB}^{(1)} \sim \tau_P^{-1/(p+1)},\quad N_{SB}^{(2)} \propto \xi \sim \tau_P^{-1/p}\, , 
\end{equation}
and  $N_{SB}^{(2)}-N_{SB}^{(1)} \sim \tau_P^{-1/p}$. Our findings show that, contrary to the transport in the  absence of BVPs, transport in presence of BVPs (at weak coupling) crucially depends on the order of the EP. This is evident from the superballistic scaling behavior where the exponent of the superballistic scaling depends on the order of the EP [Eq.~\eqref{eq:sbp}] as well as from the dependence of the phase coherence length $\xi$ on $\tau_P$ [Eq.~\eqref{eq:nsb12}].
\\


\section{Summary and outlook}
\label{sec:summary}

Transfer matrix EPs occurring at band edges of one-dimensional chains have been previously shown to lead to counterintuitive subdiffusive \cite{universal_subdiffusive} and superballistic \cite{superballistic1} scaling of conductance with system length. In this work, we have explored the effect of the order of the transfer matrix EP on such counterintuitive transport behavior at the band edges. To do so, we have  considered a one-dimensional model with hopping up to a finite range $n$ which leads to a $2n \times 2n$ non-Hermitian transfer matrix. Depending on the hopping parameters and the range of hopping, there is a systematic way to generate band-edges corresponding to arbitrarily high order EPs. Using this model, we have established analytically and demonstrated numerically that, in absence of bulk dephasing (coherent transport), conductance scaling with system length is completely unaffected by the order of the transfer matrix EP. This means that, suprisingly, $\mathcal{G}(\mu)\sim N^{-2}$, when $\mu$ corresponds to a band edge, irrespective of the order of the transfer matrix EP, although existence of an EP is crucial to get such scaling. However, in presence of bulk dephasing, the transport is affected by the order of the transfer matrix EP. The dependence of phase coherence length on the strength of dephasing, the extent of the superballistic scaling regime, as well as the exponent of the superballistic scaling, all encode the order of the transfer matrix EP.

 Our results in this paper highlight the fundamental role that transfer matrix EPs of Hermitian systems play in quantum transport. Generating arbitrary order transfer matrix EPs enable ways control phase coherence length, and design special types of environment assisted quantum transport. Our work is also an important step forward towards clarifying the super-universality of subdiffusive scaling of conductance at band edges. 
 In future, it will be interesting to explore the properties of transfer matrix for higher dimensional systems \cite{junaid_subdiffusive}, ladders \cite{ladder1,ladder2}. The fate of bulk dephasing in presence of power-law decaying hopping \cite{dhawan2024anomalous} also remains to be explored. Finally, transfer matrix EPs, and related transport properties may be tested in realistic models of materials via application of density functional theory \cite{Brandbyge_2002,Maassen_2009,Nitzan_2003,Hjorth_2017,Kurth_2019,Smidstrup_2020}.

\section*{Acknowledgements}
M. S. acknowledges financial support through National Postdoctoral Fellowship (NPDF), SERB file no.~PDF/2020/000992. M. S. and M. K. acknowledge support of the Department of Atomic Energy, Government of India, under project no. RTI4001.
B. K. A. would also like to acknowledge funding from the National Mission on Interdisciplinary  Cyber-Physical  Systems (NM-ICPS)  of the Department of Science and Technology,  Govt.~of  India through the I-HUB  Quantum  Technology  Foundation, Pune, India. B.K.A. acknowledges CRG Grant No. CRG/2023/003377 from SERB, Government of India.
M. K. thanks the VAJRA faculty scheme (No. VJR/2019/000079) from the Science and Engineering Research Board (SERB), Department of Science and Technology, Government of India.  M.K. acknowledges support from the Infosys Foundation International Exchange Program at ICTS. A.P acknowledges funding from Seed Grant from IIT Hyderabad, Project No. SG/IITH/F331/2023-24/SG-169. A.P also acknowledges funding from Japan International Coorperation Agency (JICA) Friendship 2.0 Research Grant, and from Finnish Indian Consortia for Research and Education (FICORE). B.K. A. and M. K. also thanks Katha Ganguly for insightful discussions related to the project. 

\appendix

\section{Finite-range hopping model with $n=2$ in presence of highest order (fourth order) EP}
\label{AppendixA}

In this appendix, we show the impact of highest order EP on conductance scaling in absence (Sec.~\ref{immunity_fourth}) and in presence (Sec.~\ref{lack_immunity}) of the bulk dephasing for $n=2$. In other words, we would like to show the transport behaviour when the transfer matrix has fourth order EP.


\subsection{Immunity to order of EP in scaling of transport in absence of bulk dephasing}
\label{immunity_fourth}
We would like to analyze the scaling behavior of $|\langle 1 | g(\mu)| N \rangle|^2$ which is proportional to  the conductance scaling in absence of bulk dephasing. Using Eq.~\eqref{g1na2a}, the expression for $\langle 1 | g (\mu)| N \rangle$ reduces to,
\begin{align}
\label{g12na3a}
\langle 1 | g (\mu)| N \rangle=-\frac{1}{t_2} \left \langle 1 \vert \mathbf{A}^{-1} (\mu) \vert 2 \right \rangle.
\end{align}
The $2 \times 2$ Matrix $\mathbf{A}$, defined in Eq.~\eqref{g12na3a}, has the form,
\begin{align}
\label{A1a}
\mathbf{A(\mu)}=\begin{pmatrix}
\left\langle 3 \vert \mathbf{T}^{-N} \vert 3 \right\rangle & \left\langle 3 \vert \mathbf{T}^{-N} \vert 4 \right\rangle \\
\left\langle 4 \vert \mathbf{T}^{-N} \vert 3 \right\rangle & \left\langle 4 \vert \mathbf{T}^{-N} \vert 4 \right\rangle
\end{pmatrix}.
\end{align}
Using Eq.~\eqref{A1a} in Eq.~\eqref{g12na3a}, we can obtain $\langle 1 | g (\mu)| N \rangle$ as,
\begin{align}
\label{g12na4a}
\langle 1 | g (\mu)| N \rangle=-\frac{1}{t_2}\frac{C_{1,2}(\mu)}{\mathrm{det}[\bf{A}(\mu)]}=-\frac{1}{t_2} \frac{\left \langle 3 \vert \mathbf{T}^{-N} \vert 4 \right \rangle}{\mathrm{det}\mathbf{A(\mu)}}.
\end{align}
When the chemical potential value $\mu$ corresponds to the band edge with $k=\pi$ and $t_2=t_1/4$, $\mathbf{T}(\mu)$ has a fourth-order exceptional point. We can then write the form of $R$ using Eq.~\eqref{R} as,
\begin{align}
\label{rmat2a}
R=
\begin{pmatrix}
-1 & 3 & -3 & 1 \\
1& -2 & 1 & 0 \\
-1 & 1 & 0 & 0 \\
1 & 0 & 0 & 0
\end{pmatrix}.
\end{align}
Using Eq.~\eqref{Jele} and Eq.~\eqref{rmat2a} in Eq.~\eqref{Tmata}, we can find the matrix elements of $\mathbf{T}^{-N}(\mu)$ as,
\begin{align}
\label{Tmat3a}
 \left\langle 3 \vert \mathbf{T}^{-N} \vert 3 \right\rangle &=\left\langle 1 \vert J^{-N} \vert 1 \right\rangle + \left\langle 1 \vert J^{-N} \vert 2 \right\rangle + \left\langle 1 \vert J^{-N} \vert 3 \right\rangle \nonumber \\
 &-3 \left\langle 1 \vert J^{-N} \vert 4 \right\rangle,  \nonumber \\
 \left\langle 4 \vert \mathbf{T}^{-N} \vert 3 \right\rangle & =\left\langle 1 \vert J^{-N} \vert 2 \right\rangle + 2 \left\langle 1 \vert J^{-N} \vert 3 \right\rangle + 3 \left\langle 1 \vert J^{-N} \vert 4 \right\rangle , \nonumber \\ 
\left\langle 3 \vert \mathbf{T}^{-N} \vert 4 \right\rangle & = -\left\langle 1 \vert J^{-N} \vert 4 \right\rangle ,  \nonumber \\
\left\langle 4 \vert \mathbf{T}^{-N} \vert 4 \right\rangle & =\left\langle 1 \vert J^{-N} \vert 1 \right\rangle+\left\langle 1 \vert J^{-N} \vert 2 \right\rangle + \left\langle 1 \vert J^{-N} \vert 3 \right\rangle \nonumber \\
&+  \left\langle 1 \vert J^{-N} \vert 4 \right\rangle.
\end{align}
In all the above matrix elements of $J^{-N}$, the element $\left\langle 1 \vert J^{-N} \vert 4 \right\rangle$ has the most dominant $N$ dependence that goes as $N^3$. This follows from Eq.~\eqref{Jele}. If we plug this in matrix $\mathbf{A}(\mu)$ in Eq.~\eqref{A1a} and in Eq.~\eqref{g12na4a}, it is easy to see that the numerator has dominant $N$ dependence as $N^3$. Now, the denominator has the term $\mathrm{det} \mathbf{A}(\mu)$ which goes as $N^4$. Thus, it is clear that $\langle 1 | g (\mu)| N \rangle$ has $N^3/N^4 = 1/N$ dependence and therefore $|\langle 1 | g (\mu)| N \rangle|^2 \sim 1/N^2$. Given that, the same $1/N^2$ scaling also exist for EP of second order for $n=1$, therefore this analysis shows that independent of the order of EP, the $1/N^2$ scaling of conductance at the band edge remains robust.

\subsection{Lack of immunity to order of EP in scaling of transport in presence of bulk dephasing}
\label{lack_immunity}
In this subsection, we demonstrate the lack of immunity in presence of a fourth order EP when bulk dephasing is present. Similar to Sec.~\ref{immunity_fourth} we consider $n=2$ case. We show the system size dependence of $\mathcal{G}^{(1)}_{\rm bare}(\mu)$, given in Eq.~\eqref{term1}. To calculate this quantity, we need to know all the terms of Eq.~\eqref{glng1l}. With Eq.~\eqref{Tmat3a} in hand, we now have to know the other four matrix elements of $\mathbf{T}(\mu)$ which are given below,
\begin{align}
\label{1Tmat3}
\left\langle 2 \vert \mathbf{T}^{-\ell} \vert 3 \right\rangle &=-2 \left\langle 1 \vert J^{-\ell} \vert 2 \right\rangle + 3 \left\langle 1 \vert J^{-\ell} \vert 4 \right\rangle, \nonumber \\
\left\langle 2 \vert \mathbf{T}^{-\ell} \vert 4 \right\rangle &=- \left\langle 1 \vert J^{-\ell} \vert 3 \right\rangle +  \left\langle 1 \vert J^{-\ell} \vert 4 \right\rangle, \nonumber \\
\left\langle 3 \vert \mathbf{T}^{-(N-\ell+1)} \vert 1 \right\rangle &=\left\langle 1 \vert J^{-(N-\ell+1)} \vert 3 \right\rangle  \nonumber \\ &-  \left\langle 1 \vert J^{-(N-\ell+1)} \vert 4 \right\rangle,  \nonumber \\
\left\langle 4 \vert \mathbf{T}^{-(N-\ell+1)} \vert 1 \right\rangle &=\left\langle 1 \vert J^{-(N-\ell+1)} \vert 4 \right\rangle.
\end{align}
Using  Eq.~\eqref{Tmat3a} and Eq.~\eqref{1Tmat3}, we can write the matrix elements $\langle \ell | g| N \rangle$ and $\langle 1 | g| \ell \rangle$  as, 
\begin{align}
\label{pql}
 \langle \ell | g| N \rangle &=\frac{1}{t_2} \frac{\left\langle 2 \vert \mathbf{T}^{-\ell} \vert 3 \right\rangle \left\langle 3 \vert \mathbf{T}^{-N} \vert 4 \right\rangle -\left\langle 2 \vert \mathbf{T}^{-\ell} \vert 4 \right\rangle \left\langle 3 \vert \mathbf{T}^{-N} \vert 3 \right\rangle}{\mathrm{det}\mathbf{A}} \nonumber \\
&= \frac{1}{t_2} \frac{P(\ell,N)}{\mathrm{det}\mathbf{A}},\\ \nonumber
\langle 1 | g| \ell \rangle  &=\frac{1}{t_2}\Big[ \frac{\left\langle 3 \vert \mathbf{T}^{-(N-\ell+1)} \vert 1 \right\rangle \left\langle 4 \vert \mathbf{T}^{-N} \vert 4 \right\rangle }{\mathrm{det}\mathbf{A}} \nonumber \\
&-\frac{\left\langle 4 \vert \mathbf{T}^{-(N-\ell+1)} \vert 1 \right\rangle \left\langle 3 \vert \mathbf{T}^{-N} \vert 4 \right\rangle}{ \mathrm{det}\mathbf{A}}\Big] \nonumber \\
&= \frac{1}{t_2} \frac{Q(\ell,N)}{\mathrm{det}\mathbf{A}}.
\end{align}
Thus, finally we obtain from Eq.~\eqref{term1},
\begin{align}
\label{super1a}
\mathcal{G}^{(1)}_{\rm bare}(\mu) & \sim \frac{1}{(\mathrm{det}\mathbf{A})^2 } \sum_{\ell=1}^{N} \frac{P(\ell,N)^2 Q(\ell,N)^2}{P(\ell,N)^2+ Q(\ell,N)^2} 
\end{align}
In presence of fourth order EP, we get $\mathrm{det}\mathbf{A}(\mu) \sim N^4$ and the system size dependence of the numerator $\sim N^{11}$. Thus, we see that the term  $\mathcal{G}^{(1)}_{\rm bare}(\mu) \sim N^3$ in presence of fourth order EP. This demonstrate the remarkable sensitivity of the order of EP on the conductance scaling when bulk dephasing mechanism is present. \\

\section{Relation between phase coherence length $\xi$ and the probe coupling strength $\tau_P$}
\label{largest_eigenvalue}
In this appendix, we provide an interesting relation between the phase coherence length $\xi$ and $\tau_P$ which is crucial to estimate the end of the superballistic scaling regime $N_{SB}^{(2)}$. Recall that, in presence of bulk dephasing the NEGF is given by Eq.~\eqref{NEGF} which translates to an ``effective'' transfer matrix with $\omega$ replaced by $\omega+ i \tau_p/2$. As a result, one needs to solve 
a general dispersion relation given as,
\begin{align}
\label{dispersion_like}
\omega+ i \frac{\tau_P}{2} = \epsilon (z).
\end{align}
Expanding $z$ about any real solution $k$, we can write,
\begin{align}
\label{dispersion_like1}
\omega+ i \frac{\tau_P}{2} = \epsilon (k)+ \sum_{r=1}^{\infty} \frac{(z-k)^r}{r !} \frac{\partial^r \epsilon(z)}{\partial z^r}\Big{|}_{z=k}.
\end{align}
In presence of $p$-th order EP of transfer matrix and putting $\omega=\epsilon(k)$, we can again write Eq.~\eqref{dispersion_like1} to find out the $\tau_P$ dependence of largest eigenvalue as,
\begin{align}
\label{dispersion_like2}
\frac{e^{i \pi/2}\, \tau_P}{2} &=  \sum_{r=p}^{\infty} \frac{(z-k)^r}{r !} \frac{\partial^r \epsilon(z)}{\partial z^r}\Big{|}_{z=k}  \nonumber\\
& = \frac{(z-k)^p}{p !} \frac{\partial^p \epsilon(z)}{\partial z^p}|_{z=k} \nonumber \\
&+\sum_{r=p+1}^{\infty} \frac{(z-k)^r}{r !} \frac{\partial^r \epsilon(z)}{\partial z^r}\Big{|}_{z=k}  \nonumber\\
& \approx \frac{(z-k)^p}{p !} \frac{\partial^p \epsilon(z)}{\partial z^p}\Big{|}_{z=k},
\end{align}
where we only keep the leading order term in $(z-k)$.
Thus, from Eq.~\eqref{dispersion_like2}, we obtain,
\begin{align}
\label{dispersion_like3}
z=k+ \Bigg(\frac{p! ~e^{i \pi/2} \,\tau_P}{2 \frac{\partial^p \epsilon(z)}{\partial z^p}|_{z=k}} \Bigg)^{1/p}. 
\end{align}
Therefore, the largest eigenvalue of the ``effective'' transfer matrix is of the form $\lambda_{+} = e^{\kappa_1 + i \kappa_2}$ with $k_1>0$. Then the phase coherence length is given as $\xi = |\log[\lambda_{+}]| \sim \tau_P^{1/p}$. This shows that the end of the superballistic scaling regime $N_{SB}^{(2)}$ is remarkably sensitive to the order of the EP. 


\bibliography{reference}
\end{document}